\documentstyle[12pt,epsf]{article}

\newcommand{\beq}{\begin{equation}}
\newcommand{\eeq}{\end{equation}}
\newcommand{\bea}{\begin{eqnarray}}
\newcommand{\eea}{\end{eqnarray}}

\renewcommand{\a}{\alpha}
\newcommand{\qt}{q_\perp}

\newcommand{\pta}{p_{a'\perp}}
\newcommand{\ptb}{p_{b'\perp}}
\newcommand{\ra}{\rightarrow}
\newcommand{\lra}{\leftrightarrow}
\newcommand{\s}{\hat s}
\newcommand{\th}{\hat t}
\newcommand{\nn}{\nonumber}
\newcommand{\un}{\underline}

\begin{document}
\topskip 2cm 
\begin{titlepage}

\hspace*{\fill}\parbox[t]{4cm}{EDINBURGH 96/4\\ May 1996}

\vspace{2cm}

\begin{center}
{\large\bf Next-to-Leading Corrections to the BFKL Equation} \\
\vspace{1.5cm}
{\large Vittorio Del Duca} \\
\vspace{.5cm}
{\sl Particle Physics Theory Group,\,
Dept. of Physics and Astronomy\\ University of Edinburgh,\,
Edinburgh EH9 3JZ, Scotland, UK}\\
\vspace{1.5cm}
\vfil
\begin{abstract}

Using the helicity formalism in the high-energy limit, we compute
the amplitudes which generate the real next-to-leading-logarithmic 
corrections to the BFKL equation. Accordingly, we provide a 
list of all the off-shell vertices necessary to build such amplitudes.

\end{abstract}

\vspace{2cm}

{\sl To appear in the Proceedings of\\ Les Rencontres de Physique de la
Vallee d'Aoste\\ La Thuile, Val d'Aosta, Italy, March 1996}
\end{center}

\end{titlepage}

\section{Introduction}
\label{sec:uno}

Semi-hard strong-interaction processes, for which the squared
center-of-mass energy $s$ is much larger than the momentum transfer
$Q^2$, have attracted a lot of interest in the latest years, because
of the large kinematic region explored in deeply inelastic scattering (DIS) by 
the electron-proton collider at HERA, where values of $x_{bj}=Q^2/s$ of the
order of $10^{-5}$ have been attained \cite{hera}. The evolution of the 
$F_2(x_{bj},Q^2)$ structure function in $\ln(Q^2)$ is described by the 
Dokshitzer-Gribov-Lipatov-Altarelli-Parisi (DGLAP) equation to an accuracy
determined by the order of $\a_s$ to which we compute the expansion of
the splitting functions $P_{ab}(x,\a_s)$, with $a,\,b=$ quarks or gluons.
At very small values of $x_{bj}$ we may consider to resum
the leading (logarithmic) contributions in $1/x$ to the 
splitting functions to all orders in $\a_s$.
This may be performed by using the Balitsky-Fadin-Kuraev-Lipatov (BFKL) 
evolution equation \cite{bal}, which allows us to compute the gluon splitting
functions at leading logarithmic (LL) accuracy in $1/x$ \cite{bcm}, and the 
quark splitting functions at next-to-leading logarithmic (NLL) accuracy 
\cite{cat}. In order to know the gluon splitting functions at NLL accuracy, 
the NLL corrections to the BFKL equation must be computed. 

However, a caveat is in order: the BFKL equation computes the
radiative corrections to parton-parton scattering in the high-energy limit,
assuming that the outgoing partons are balanced in transverse momentum.
Therefore only one hard transverse-momentum scale is allowed in the
process. It is not possible to assess whether this is realized in the 
configurations that
drive the rise of $F_2$ at small $x_{bj}$. Such a constraint, though, may be 
forced upon the DIS process at small $x_{bj}$ by tagging a jet in the 
proton direction and by requiring that the squared jet transverse momentum is 
of the order of $Q^2$ \cite{muel}. An analogous process for which one may
consider to resum the leading logarithms, in $\ln(\s/Q^2)$, in the
partonic cross section is two-jet production at large rapidity intervals 
$\Delta\eta$
\cite{mn}, a process which may be explored at the Tevatron $\bar p\, p$
collider. A comparison of the ${\cal O}(\a_s^3)$ matrix elements, exact and
in the multi-Regge approximation used in the BFKL calculation, shows though
that the discrepancies are quite big at the not very large values of
$\Delta\eta$ attainable at the Tevatron, and would still be sizeable at
the LHC collider \cite{DDS}. This gives one more reason to compute the
NLL corrections to the BFKL equation.

The working tools of the BFKL theory are the 
Fadin-Kuraev-Lipatov (FKL) multigluon amplitudes 
\cite{lip}, \cite{FKL} in the multi-Regge kinematics,
which requires that the final-state partons are strongly ordered in
rapidity and have comparable transverse momentum.
The parton-parton scattering may be initiated by
either quarks or gluons, however in the high-energy limit the leading
contribution comes only from gluon exchange in the cross channel, therefore 
the leading corrections to parton-parton scattering are purely gluonic. The 
building blocks of the tree-level FKL amplitudes are the process-dependent
helicity-conserving vertices $g^*\, g \rightarrow g$, eq.(\ref{centrc}) 
with $g^*$ an off-shell gluon, and $g^*\, q \rightarrow q$, or $g^*\,\bar q 
\rightarrow \bar q$, eq.(\ref{cbqqm}) and (\ref{cbqqp}), which produce a 
parton at either end of the ladder; and the process-independent Lipatov 
vertex $g^*\, g^* \rightarrow g$, eq.(\ref{lip}), which emits a gluon along 
the ladder. The helicity-conserving and the Lipatov vertices, and accordingly
the FKL amplitudes, assume a simpler analytic form when the helicity of 
the produced gluons is explicitly fixed \cite{lipat}, \cite{ptlip}. 
The LL virtual radiative corrections then {\sl reggeize} the gluons, i.e.
make the gluon propagators exchanged in the cross channel to assume a 
Regge-like form, eq.(\ref{sud}). The Lipatov vertex and the
reggeized gluon enter the BFKL equation, and the helicity-conserving
vertices fix the boundary conditions to it.

The NLL corrections to the FKL amplitudes are divided into
real corrections, induced by the corrections to the multi-Regge kinematics
\cite{fl}-\cite{ptlipqq}, and virtual NLL corrections.
The real corrections to the tree-level FKL amplitudes arise from the 
kinematical regions in which two partons are produced with
similar rapidity, either at the ends of or along the ladder, termed
the forward-rapidity and the central-rapidity regions respectively.
The building blocks of these amplitudes are the vertices which
describe the emission of two partons
in the forward-rapidity region, $g^*\, g \rightarrow g\, g$ eq.(\ref{nllfg}),
$g^*\, g \rightarrow \bar{q}\, q$ eq.(\ref{forwqq}), and $g^*\, q 
\rightarrow g\, q$ eq.(\ref{forwqg}) and (\ref{forwqgb}); and
in the central-rapidity region,
$g^*\, g^* \rightarrow g\, g$ eq.(\ref{centrb}) or $g^*\, g^* \rightarrow 
\bar{q}\, q$ eq.(\ref{centrq}).
The vertices for the emission in the central-rapidity region determine
the real NLL corrections to the BFKL equation \cite{fl2}, and the
vertices for the emission in the forward-rapidity region fix the boundary
conditions to it. All vertices transform into their complex conjugates under
helicity reversal.

\section{Tree-level amplitudes in the helicity formalism}
\label{sec:due}

A tree-level multigluon amplitude in a helicity basis has
the form \cite{mp}
\begin{equation}
M_n = \sum_{[a,1,...,n,b]'} {\rm tr}(\lambda^a\lambda^{d_1} \cdots
\lambda^{d_n} \lambda^b) \, m(-p_a,-\nu_a; p_1,\nu_1;...;
p_n,\nu_n; -p_b,-\nu_b)\, ,\label{one}
\end{equation}
where $a,d_1,..., d_n,b$, and $\nu_a,\nu_1,...,\nu_b$
are respectively the colors and the helicities of the gluons,
the $\lambda$'s are the color matrices in the fundamental representation
of SU($\rm N_c$), 
the sum is over the noncyclic permutations of the color orderings 
$[a,1,...,b]$ and all the momenta are taken as outgoing. 
For the {\sl maximally helicity-violating}
configurations $(-,-,+,...,+)$, the subamplitudes
$m(-p_a,-\nu_a; p_1,\nu_1;...; p_n,\nu_n; -p_b,-\nu_b)$, invariant with
respect to tranformations between physical gauges,
assume the form \cite{pt},
\begin {equation}
m(-,-,+,...,+) = 2^{1+n/2}\, g^n\, {\langle p_i p_j\rangle^4\over
\langle p_a p_1\rangle \cdots\langle p_n p_b\rangle 
\langle p_b p_a\rangle}\, ,\label{two}
\end{equation}
with $i$ and $j$ the gluons of negative helicity, and with 
the spinor products defined as
\begin{eqnarray}
\langle p k\rangle &=& \langle p- | k+ \rangle = \overline{\psi_-(p)}
\psi_+(k)\, ,\label{cpro}\\ 
\left[pk\right] &=& \langle p+ | k- \rangle = \overline{\psi_+(p)}\psi_-(k)\, 
,\nonumber
\end{eqnarray}
through massless Dirac spinors of fixed helicity, $\psi_{\pm}(p)$.
The subamplitudes
(\ref{two}) are {\sl exact}, and in computing them the representation 
\begin {equation}
\epsilon_{\mu}^{\pm}(p,k) = \pm {\langle p\pm |\gamma_{\mu}| k\pm\rangle\over
\sqrt{2} \langle k\mp | p\pm \rangle}\, ,\label{hpol}
\end{equation}
for the gluon polarization has been used, with $k$ an arbitrary light-like
momentum. The ordering of the spinor products in the denominator of 
eq.(\ref{two}) is set by the permutation of
the color ordering $[a,1,...,b]$. The configurations
$(+,+,-,...,-)$ are then obtained by replacing the $\langle p k\rangle$
products with $\left[k p\right]$ products. 

A tree-level multigluon amplitude with a quark-antiquark pair
has the form \cite{mp},
\begin{equation}
M_n = \sum_{[1,...,n]} (\lambda^{d_1} \cdots \lambda^{d_n})_{i\bar{j}} \, 
m(q,\nu; p_1,\nu_1;...; p_n,\nu_n; \bar{q},-\nu)\, ,\label{due}
\end{equation}
where ($i,\bar{j}$) are the color indices of the quark-antiquark pair,
the sum is over the permutations of the color orderings $[1,...,n]$,
and we have taken into account that helicity is conserved over the
quark line.
For the {\sl maximally helicity-violating} configurations, $(-,-,+,...,+)$,
the subamplitudes are
\begin {eqnarray}
m(\bar{q}^+; q^-; g_1;...; g_n) &=& 2^{n/2}\, g^n\, 
{\langle \bar{q} p_i \rangle \langle q p_i \rangle^3 \over
\langle \bar{q} q\rangle \langle q p_1\rangle \cdots\langle p_n 
\bar{q}\rangle} \label{dueb}\\
m(\bar{q}^-; q^+; g_1;...; g_n) &=& 2^{n/2}\, g^n\, 
{\langle \bar{q} p_i \rangle^3 \langle q p_i \rangle \over
\langle \bar{q} q\rangle \langle q p_1\rangle \cdots\langle p_n 
\bar{q}\rangle}\, ,\nn
\end{eqnarray}
where the $i^{th}$ gluon has negative helicity, and the ordering
of the spinor products in the denominator is set by the permutation of
the color ordering $[1,...,n]$. The subamplitudes
(\ref{two}) and (\ref{dueb}) are related by a supersymmetric
Ward identity \cite{mp}.

\section{Amplitudes in the multi-Regge kinematics}
\label{sec:tre}

\begin{figure}[htb]
\vspace*{-6cm}
\hspace*{-1cm}
\epsfxsize=18cm \epsfbox{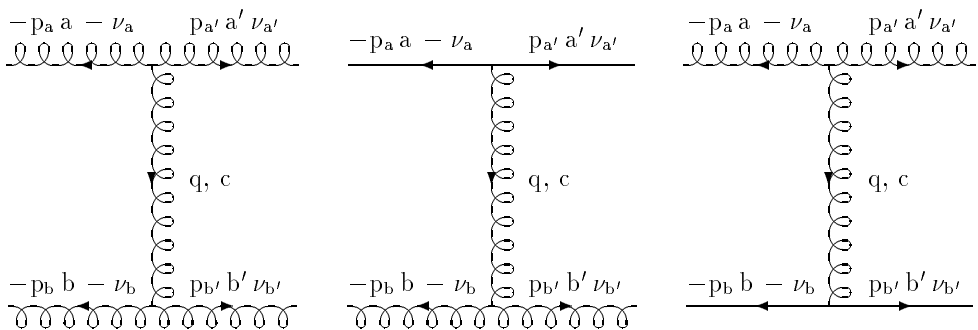}
\vspace*{-15.5cm}
\caption{$(a)$ Amplitude for $g\, g \ra g\, g$ scattering and $(b),\,(c)$ for
$q\, g \ra q\, g$ scattering. We label the external lines with momentum, 
always taken as outgoing, color and helicity,
and the internal lines with momentum and color.}
\label{fig:one}
\end{figure}

We consider the elastic scattering of two gluons of momenta $p_a$ and $p_b$
in two gluons of momenta $p_{a'}$ and $p_{b'}$, in the high-energy limit
$\s\gg |\th|$ (Fig.~\ref{fig:one}a). Let $y$ be the rapidity difference 
between the outgoing
gluons, and $\ptb = -\pta = \qt$. Then $\th \simeq -|\qt|^2$, and
$\s \simeq -\th e^y$, and the high-energy limit implies that $y\gg 1$.
Using eq.(\ref{two}) and the spinor products in the representation of 
ref.~\cite{ptlip}, and rewriting the traces of $\lambda$ matrices
as products of structure constants,
\beq
[\lambda^a,\lambda^b] = i\, f^{abc}\, \lambda^c\, ,\qquad
{\rm tr} (\lambda^a \lambda^b) = {\delta_{ab}\over 2}\, ,\label{alg}
\eeq
the scattering amplitude in the high-energy limit may be written as,
\beq
M^{aa'bb'}_{\nu_a\nu_{a'}\nu_{b'}\nu_b} = 2 {\hat s}
\left[i g\, f^{aa'c}\, C^{gg}_{-\nu_a\nu_{a'}}(-p_a,p_{a'}) \right]
{1\over \th} \left[i g\, f^{bb'c}\, C^{gg}_{-\nu_b\nu_{b'}}(-p_b,p_{b'}) 
\right]\, ,\label{elas}
\eeq
with the helicity-conserving vertices $g^*\, g \rightarrow g$, 
with $g^*$ an off-shell gluon, \cite{FKL}, \cite{ptlip}
\beq
C_{-+}^{gg}(-p_a,p_{a'}) = 1 \qquad C_{-+}^{gg}(-p_b,p_{b'}) =
{p_{b'\perp}^* \over p_{b'\perp}}\, ,\label{centrc}
\eeq
where we use the complex notation, $p_\perp = p_x + i p_y$, for the
transverse momentum. The $C$-vertices transform
into their complex conjugates under helicity reversal,
$C_{\{\nu\}}^*(\{k\}) = C_{\{-\nu\}}(\{k\})$. The helicity-flip
vertex $C_{++}$ is subleading in the high-energy limit.
Using eq.(\ref{due}-\ref{alg}), the quark-gluon $q\, g \rightarrow q\, g$
scattering amplitude in the high-energy limit is,
\bea
M^{q\, g \rightarrow q\, g} &=& 2 {\hat s} \left[g\, \lambda^c_{a' \bar a}\,
C_{-\nu_a \nu_a}^{\bar q q}(-p_a,p_{a'}) \right] {1\over \th} 
\left[i g\, f^{bb'c}\, C^{gg}_{-\nu_b\nu_{b'}}(-p_b,p_{b'}) \right]\, 
,\label{elasqa} \\
M^{g\, q \rightarrow g\, q} &=& 2 {\hat s} \left[i g\, f^{aa'c}\, 
C^{gg}_{-\nu_a\nu_{a'}}(-p_a,p_{a'}) \right] {1\over \th}
\left[g\, \lambda^c_{b' \bar b}\,
C_{-\nu_b \nu_b}^{\bar q q}(-p_b,p_{b'}) \right]\, ,\label{elasqb}
\eea
where the antiquark is $-p_a$ in eq.(\ref{elasqa}) (Fig.~\ref{fig:one}b)
and $-p_b$ in eq.(\ref{elasqb}) (Fig.~\ref{fig:one}c), 
and the $C$-vertices $g^*\, q \rightarrow q$ are,
\beq
C_{-+}^{\bar q q}(-p_a,p_{a'}) = 1\, ;\qquad C_{-+}^{\bar q q}(-p_b,p_{b'}) =
\left({p_{b'\perp}^* \over p_{b'\perp}}\right)^{1/2}\, .\label{cbqqm}
\eeq
Analogously, the antiquark-gluon $\bar q\, g \rightarrow \bar q\, g$
scattering amplitude is,
\bea
M^{\bar q\,g\rightarrow\bar q\,g} &=& 2 {\hat s} \left[g\,\lambda^c_{a\bar a'}
\, C_{-\nu_a \nu_a}^{q \bar q}(-p_a,p_{a'}) \right] {1\over \th} 
\left[i g\, f^{bb'c}\, C^{gg}_{-\nu_b\nu_{b'}}(-p_b,p_{b'}) \right]\, 
,\label{elasaqa}\\ 
M^{g\,\bar q\rightarrow g\,\bar q} &=& 2 {\hat s} \left[i g\, f^{aa'c}\, 
C^{gg}_{-\nu_a\nu_{a'}}(-p_a,p_{a'}) \right] {1\over \th}
\left[g\, \lambda^c_{b \bar b'}\,
C_{-\nu_b \nu_b}^{q \bar q}(-p_b,p_{b'}) \right]\, ,\label{elasaqb}
\eea
where the antiquark is $p_{a'}$ in eq.(\ref{elasaqa}) (Fig.~\ref{fig:one}b) 
and $p_{b'}$ in eq.(\ref{elasaqb}) (Fig.~\ref{fig:one}c), and the 
$C$-vertices $g^*\, \bar q \rightarrow \bar q$ are,
\beq
C_{-+}^{q \bar q}(-p_a,p_{a'}) = -1\, ;\qquad C_{-+}^{q \bar q}(-p_b,p_{b'}) =
- \left({p_{b'\perp}^* \over p_{b'\perp}}\right)^{1/2}\, .\label{cbqqp}
\eeq
In the amplitudes (\ref{elas}), (\ref{elasqa}), (\ref{elasqb}),
(\ref{elasaqa}), (\ref{elasaqb}),
the leading contributions from all the 
Feynman diagrams have been included. However, the amplitudes have the
effective form of a gluon exchange in the $t$ channel (Fig.~\ref{fig:one}),
and differ only for the relative color strength in the production vertices
\cite{CM}.
This allows us to replace an incoming gluon with a quark, for instance
on the upper line, via the simple substitution
\beq
i g\, f^{aa'c}\, C^{gg}_{-\nu_a\nu_{a'}}(-p_a,p_{a'}) \lra g\, 
\lambda^c_{a' \bar a}\, C_{-\nu_a \nu_a}^{\bar q q}(-p_a,p_{a'})\, 
,\label{qlrag}
\eeq
and similar ones for an antiquark and/or for the lower line.

\begin{figure}[htb]
\vspace*{-5.5cm}
\hspace*{-2cm}
\epsfxsize=18cm \epsfbox{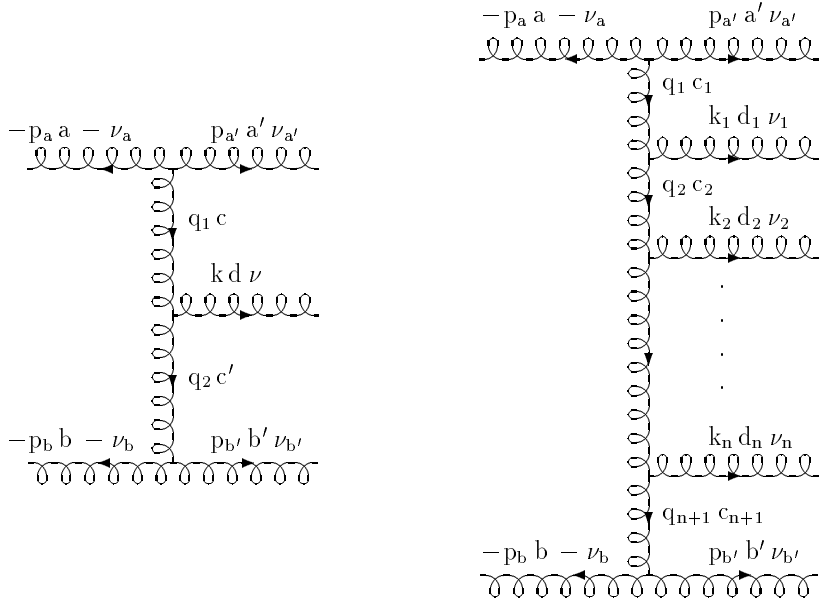}
\vspace*{-12.5cm}
\caption{$(a)$ Amplitude for $g\, g \ra g\, g\, g$ scattering, and $(b)$ 
for the production of $n+2$ gluons.}
\label{fig:due}
\end{figure}

Next, we consider the production of three gluons of momenta $p_{a'}$,
$k$ and $p_{b'}$ (Fig.~\ref{fig:due}a), and we require that the gluons are 
strongly ordered in their rapidities and have comparable transverse momenta,
\begin{equation}
y_{a'} \gg y \gg y_{b'};\qquad |p_{a'\perp}|\simeq|k_\perp|\simeq|\ptb|\, 
.\label{treg}
\end{equation}
Eq.(\ref{treg}) is the simplest example of {\sl multi-Regge kinematics}.
Using eq.(\ref{two}) and the algebra (\ref{alg}), the scattering amplitude is,
\begin{eqnarray}
M^{g g \ra g g g} &=& 
2 {\hat s} \left[i g\, f^{aa'c}\, C_{-\nu_a\nu_{a'}}^{gg}(-p_a,p_{a'})
\right]\, {1\over\hat t_1}\, \label{three}\\ &\times& \left[i g\, f^{cdc'}\, 
C^g_{\nu}(q_1,q_2)\right]\, {1\over \hat t_2}\, 
\left[i g\, f^{bb'c'}\, C_{-\nu_b\nu_{b'}}^{gg}(-p_b,p_{b'}) \right]\, ,\nn
\end{eqnarray}
with $p_{a'\perp} = - q_{1\perp}$, $p_{b'\perp} = q_{2\perp}$ and
$\th_i \simeq - |q_{i\perp}|^2$ with $i=1,2$ and with the Lipatov vertex
$g^*\, g^* \rightarrow g$ \cite{lip}, \cite{lipat}, \cite{ptlip},
\beq
C^g_+(q_1,q_2) = \sqrt{2}\, {q^*_{1\perp} q_{2\perp}\over k_\perp}\, 
.\label{lip}
\eeq
The amplitude (\ref{three}) has the effective form of a gluon-ladder 
exchange in the $t$ channel, however the additional gluon $k$ has been 
inserted either along the ladder (Fig.~\ref{fig:due}a) or as a 
bremsstrahlung gluon on the external legs. In this sense the Lipatov
vertex (\ref{lip}) is a non-local effective vertex.
Again, we may replace an incoming gluon with a quark
via the substitution (\ref{qlrag}).

Eq.(\ref{three}) generalizes to the production of $n+2$ gluons \cite{FKL} 
(Fig.~\ref{fig:due}b) in the multi-Regge kinematics,
\begin{equation}
y_{a'} \gg y_1 \gg ...\gg y_n \gg y_{b'};\qquad 
|p_{i\perp}|\simeq|p_{\perp}|\, ,\label{mreg}
\end{equation}
with $i=a',1,...,n,b'$ in a straightforward manner,
\begin{eqnarray}
M^{aa'd_1...d_nb'b}_{\nu_a\nu_{a'}\nu_1...\nu_n\nu_{b'}\nu_b} &=& 2 {\hat s}
\left[i g\, f^{aa'c_1}\, C_{-\nu_a\nu_{a'}}^{gg}(-p_a,p_{a'}) \right]\,
{1\over\hat t_1}
\nonumber\\ &\times& \left[i g\, f^{c_1d_1c_2}\, C^g_{\nu_1}(q_1,q_2)\right]\,
{1\over \hat t_2} \nonumber\\ &\times&
\label{ntree}\\ &\times&\nonumber\\ &\times& \left[i g\, f^{c_nd_nc_{n+1}}\, 
C^g_{\nu_n}(q_n,q_{n+1})\right]\, 
{1\over \hat t_{n+1}} \nonumber\\ &\times& \left[i g\, f^{bb'c_{n+1}}\, 
C_{-\nu_b\nu_{b'}}^{gg}(-p_b,p_{b'}) \right]\, ,\nonumber
\end{eqnarray}
with $\th_i \simeq - |q_{i\perp}|^2$ and $i=1,...,n+1$. Thus the tree-level
FKL amplitudes are built by repeatedly using the process-independent 
Lipatov vertex (\ref{lip}) for the gluon emission along the ladder, and 
are bounded by process-dependent outer vertices, which for gluons are given
by eq.(\ref{centrc}) and for (anti)quarks by eq.(\ref{cbqqm}) and 
(\ref{cbqqp}). However, no quarks may be produced along the ladder
since that would involve quark exchange in the $t$ channel, which is
suppressed in the kinematics (\ref{mreg}).

%
%
%
%

\section{Amplitudes in next-to-leading approximation}
\label{sec:four}

In order to obtain amplitudes in next-to-leading approximation,
we must relax the strong rapidity ordering between the produced gluons
(\ref{mreg}), and allow for the production of two gluons or of a $q\bar q$
pair with similar rapidity. 

\subsection{The forward-rapidity region}
\label{sec:fourone}

\begin{figure}[htb]
\vspace*{-4.5cm}
\hspace*{-2cm}
\epsfxsize=18cm \epsfbox{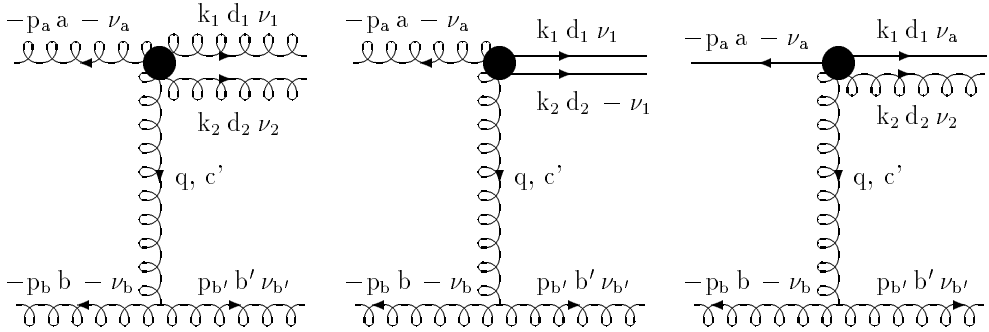}
\vspace*{-17cm}
\caption{Amplitudes for the production of three partons, with partons
$k_1$ and $k_2$ in the forward-rapidity region of parton $k_0$.}
\label{fig:forw}
\end{figure}

We begin with the simplest case, i.e. 
the production of three partons of momenta $k_1$, $k_2$ and $p_{b'}$ 
in the scattering between two partons of momenta $p_a$ and $p_b$, 
with partons $k_1$ and $k_2$ in the forward-rapidity
region of parton $p_a$,
\begin{equation}
y_1 \simeq y_2 \gg y_{b'}\,;\qquad |k_{1\perp}|\simeq|k_{2\perp}|
\simeq|p_{b'\perp}|\, .\label{qmreg}
\end{equation}
First we consider the amplitude for the scattering $g\, g\, \ra g\, g\, g$
(Fig.~\ref{fig:forw}a) \cite{fl}, \cite{ptlipnl}. Using eq.(\ref{two}) we
obtain
\begin{eqnarray}
& & M^{gg}(-p_a,-\nu_a; k_1,\nu_1; k_2,\nu_2; p_{b'},\nu_{b'}; -p_b,-\nu_b)
\nonumber\\ & & = 2\sqrt{2}\, g^3\, {\hat s\over |p_{b'\perp}|^2}\, 
C^{g\,g}_{-\nu_b\nu_{b'}}(-p_b,p_{b'})\, C^{g\,g\,g}_{-\nu_a\nu_1\nu_2}
(-p_a,k_1,k_2) 
\left\{ A_{\Sigma\nu_i}(-p_a,k_1,k_2) \right. \label{trepos}\\ & & \times
{\rm tr} \left( \lambda^a \lambda^{d_1} \lambda^{d_2} \lambda^{b'} 
\lambda^b - \lambda^a \lambda^{d_1} \lambda^{d_2} \lambda^b \lambda^{b'} 
+ \lambda^a \lambda^{b'} \lambda^b \lambda^{d_2} \lambda^{d_1} - 
\lambda^a \lambda^b \lambda^{b'} \lambda^{d_2} \lambda^{d_1} \right) 
\nonumber\\ & &
\left. - B_{\Sigma\nu_i}(-p_a,k_1,k_2)\, {\rm tr} \left(
\lambda^a \lambda^{d_1} \lambda^{b'} \lambda^b \lambda^{d_2}
- \lambda^a \lambda^{d_2} \lambda^b \lambda^{b'} \lambda^{d_1} \right) 
+ \left(\begin{array}{c} k_1\leftrightarrow k_2\\ d_1\leftrightarrow d_2 
\end{array}\right)\right\}\, ,\nonumber
\end{eqnarray}
with the vertex $C^{g\,g}_{-\nu_b\nu_{b'}}(-p_b,p_{b'})$ as in 
eq.(\ref{centrc}), $\sum\nu_i=-\nu_a+\nu_1+\nu_2$ and,
\begin{eqnarray}
C^{g\,g\,g}_{-++}(-p_a,k_1,k_2) = 1\, ; & & C^{g\,g\,g}_{+-+}(-p_a,k_1,k_2) 
= {1 \over\left(1+{k_2^+\over k_1^+}\right)^2}\, ; \label{treposb}\\ 
C^{g\,g\,g}_{++-}(-p_a,k_1,k_2) = {1 \over\left(1+{k_1^+\over k_2^+} 
\right)^2}\, ; & & A_+(-p_a,k_1,k_2) = 2\, {p_{b'\perp}\over k_{1\perp}}
{1\over k_{2\perp} - k_{1\perp} {k_2^+\over k_1^+}}\, ; \nn\\ 
B_{\Sigma\nu_i}(-p_a,k_1,k_2) &=& A_{\Sigma\nu_i}(-p_a,k_1,k_2) + 
A_{\Sigma\nu_i}(-p_a,k_2,k_1)\, ,\label{ab}
\end{eqnarray}
with the production vertex of gluons $k_1$ and $k_2$ given by the 
product of the vertex $C^{g\,g\,g}(-p_a,k_1,k_2)$ with either $A$ or $B$,
where we have used light-cone coordinates $k^{\pm}= k_0\pm k_z$. The vertex 
$C^{g\,g\,g}_{+++}(-p_a,k_1,k_2)$ is subleading to the required accuracy.
The vertex $A_{\Sigma\nu_i}$ has a collinear divergence as $2k_1\cdot 
k_2\ra 0$, but the divergence cancels out in the vertex $B_{\Sigma\nu_i}$
where gluons 1 and 2 are not adjacent in color ordering \cite{ptlipnl}.
Using the algebra (\ref{alg}) and eq.(\ref{ab}), and fixing $\th \simeq
- |p_{b'\perp}|^2$, the amplitude (\ref{trepos}) may be rewritten as,
\begin{eqnarray}
& & M^{gg}(-p_a,-\nu_a; k_1,\nu_1; k_2,\nu_2; p_{b'},\nu_{b'}; -p_b,-\nu_b)
\label{nllfg}\\ & & = 2\hat s \left\{ C^{g\,g\,g}_{-\nu_a\nu_1\nu_2}
(-p_a,k_1,k_2) \left[ (ig)^2\, f^{ad_1c} f^{cd_2c'} {1\over\sqrt{2}}\, 
A_{\Sigma\nu_i}(-p_a,k_1,k_2) + \left(\begin{array}{c} k_1\leftrightarrow k_2\\
d_1\leftrightarrow d_2 \end{array}\right) \right] \right\} \nn\\ & & \times
{1\over \th}\, \left[ig\, f^{bb'c'} C^{g\,g}_{-\nu_b\nu_{b'}}(-p_b,p_{b'})
\right]\, ,\nn
\end{eqnarray}
where we have enclosed the production vertex $g^*\, g \rightarrow g\, g$
of gluons $k_1$ and $k_2$ 
in curly brackets. In the multi-Regge limit $k_1^+\gg k_2^+$ it becomes
\begin{eqnarray}
\lim_{k_1^+\gg k_2^+} {1\over\sqrt 2} C^{g\,g\,g}_{-\nu_a\nu_1\nu_2}
(-p_a,k_1,k_2) A_{\Sigma\nu_i}(-p_a,k_1,k_2) = C^{g\,g}_{-\nu_a\nu_1}(-p_a,k_1)
{1\over \th_{12}}\, C^g_{\nu_2}(q_{12},q)\, ,\label{camr}
\end{eqnarray}
with $q_{12}$ the momentum of the gluon exchanged between $k_1$ and
$k_2$ in the multi-Regge limit, and $\th_{12} \simeq - |q_{12\perp}|^2$;
thus the amplitude (\ref{nllfg}) reduces to eq.(\ref{three}), as expected.

Using eq.(\ref{due}-\ref{alg}),
the amplitude for the production of a $q\bar q$ pair in the forward-rapidity
region of gluon $p_a$ is (Fig.~\ref{fig:forw}b), \cite{fk}, \cite{ptlipqq},
\begin{eqnarray}
& & M^{\bar{q}q}(-p_a,-\nu_a; k_1,\nu_1; k_2,-\nu_1; p_{b'},\nu_{b'}; 
-p_b,-\nu_b) \label{forwqq}\\ & & = 2\hat s \left\{\sqrt{2}\, g^2\, 
C^{g\,\bar{q}\,q}_{-\nu_a\nu_1-\nu_1}(-p_a,k_1,k_2) \left[\left(\lambda^{c'} 
\lambda^a\right)_{d_2\bar{d_1}} A_{-\nu_a}(k_1,k_2) + \left(\lambda^a 
\lambda^{c'}\right)_{d_2\bar{d_1}} A_{-\nu_a}(k_2,k_1) \right] \right\} 
\nonumber\\ & & \times {1\over \th}\, \left[ig\, f^{bb'c'} 
C^{g\,g}_{-\nu_b\nu_{b'}}(-p_b,p_{b'}) \right]\, ,\nn
\end{eqnarray}
with $k_1$ the antiquark, the production vertex $g^*\, g 
\rightarrow \bar q q$ in curly brackets, $A$ defined in eq.(\ref{treposb}),
and $C^{g\,\bar{q}\,q}$ given by,
\begin{eqnarray}
C^{g\,\bar{q}\,q}_{++-}(-p_a,k_1,k_2) &=& {1 \over 2} \sqrt{k_1^+\over k_2^+}
{1 \over\left(1+{k_1^+\over k_2^+} \right)^2} \label{cqqa}\\
C^{g\,\bar{q}\,q}_{+-+}(-p_a,k_1,k_2) &=& {1 \over 2} \sqrt{k_2^+\over k_1^+}
{1 \over\left(1+{k_2^+\over k_1^+} \right)^2}\, .\nn
\end{eqnarray}
Note that the vertex $B_{\Sigma\nu_i}$ (\ref{ab}) does not appear in 
eq.(\ref{forwqq}) because the $\bar{q}\,q$ pair is bound to be adjacent 
in color ordering [cf. eq.(\ref{due})]. In the multi-Regge limit 
$k_1^+\gg k_2^+$ the vertices (\ref{cqqa}) vanish, in agreement with the
remark at the end of sect.~3.

Using again eq.(\ref{due}-\ref{alg}),
the amplitude for the production of a $q\, g$ pair in the forward-rapidity
region of quark $p_a$ is (Fig.~\ref{fig:forw}c),
\begin{eqnarray}
& & M^{\bar{q}q}(-p_a,-\nu_a; k_1,\nu_a; k_2,\nu_2; p_{b'},\nu_{b'}; 
-p_b,-\nu_b) \label{forwqg}\\ & & = 2\hat s \left\{\sqrt{2}\, g^2\, 
C^{\bar{q}\,q\,g}_{-\nu_a\nu_a\nu_2}(-p_a,k_1,k_2) \left[\left(\lambda^{d_2}
\lambda^{c'}\right)_{d_1\bar{a}} A_{\nu_2}(k_1,k_2) - \left(\lambda^{c'}
\lambda^{d_2}\right)_{d_1\bar{a}} B_{\nu_2}(k_1,k_2)\right] \right\} 
\nonumber\\ & & \times {1\over \th}\, \left[ig\, f^{bb'c'} 
C^{g\,g}_{-\nu_b\nu_{b'}}(-p_b,p_{b'}) \right]\, ,\nn
\end{eqnarray}
with $-p_a$ the antiquark, the production vertex $q\, g^* \rightarrow q\, g$ 
in curly brackets, $A$ and $B$ defined in eq.(\ref{treposb}) and (\ref{ab}), 
and
\begin{eqnarray}
C^{\bar{q}\,q\,g}_{-++}(-p_a,k_1,k_2) &=& {1 \over 2} 
{1 \over\left(1+{k_2^+\over k_1^+} \right)^{1/2}}\, ,\label{cqg}\\
C^{\bar{q}\,q\,g}_{+-+}(-p_a,k_1,k_2) &=& {1 \over 2} 
{1 \over\left(1+{k_2^+\over k_1^+} \right)^{3/2}}\, .\nn
\end{eqnarray}
Analogously, when $k_1$ is the antiquark we obtain,
\begin{eqnarray}
& & M^{q\bar{q}}(-p_a,-\nu_a; k_1,\nu_a; k_2,\nu_2; p_{b'},\nu_{b'}; 
-p_b,-\nu_b) \label{forwqgb}\\ & & = -2\hat s \left\{\sqrt{2}\, g^2\, 
C^{q\,\bar{q}\,g}_{-\nu_a\nu_a\nu_2}(-p_a,k_1,k_2) \left[\left(\lambda^{c'}
\lambda^{d_2}\right)_{a\bar{d_1}} A_{\nu_2}(k_1,k_2) - \left(\lambda^{d_2}
\lambda^{c'}\right)_{a\bar{d_1}} B_{\nu_2}(k_1,k_2)\right] \right\} 
\nonumber\\ & & \times {1\over \th}\, \left[ig\, f^{bb'c'} 
C^{g\,g}_{-\nu_b\nu_{b'}}(-p_b,p_{b'}) \right]\, ,\nn
\end{eqnarray}
with
\begin{eqnarray}
C^{q\,\bar{q}\,g}_{-++}(-p_a,k_1,k_2) &=& -{1 \over 2} 
{1 \over\left(1+{k_2^+\over k_1^+} \right)^{1/2}}\, ,\label{cqgb}\\
C^{q\,\bar{q}\,g}_{+-+}(-p_a,k_1,k_2) &=& -{1 \over 2} 
{1 \over\left(1+{k_2^+\over k_1^+} \right)^{3/2}}\, .\nn
\end{eqnarray}
In the multi-Regge limit $k_1^+\gg k_2^+$ the amplitudes (\ref{forwqg})
and (\ref{forwqgb}) reduce to eq.(\ref{three}), with the substitution 
(\ref{qlrag}) for the upper line, and respectively the vertices 
$C^{\bar{q}\,q}$, eq.(\ref{cbqqm}), and $C^{q\,\bar{q}}$, eq.(\ref{cbqqp}).
Finally, the amplitudes of Fig.~\ref{fig:forw} with a quark in the lower
line are obtained via the corresponding substitution (\ref{qlrag}).

\subsection{The central-rapidity region}
\label{sec:fourtwo}

\begin{figure}[htb]
\vspace*{-5cm}
\hspace*{0cm}
\epsfxsize=18cm \epsfbox{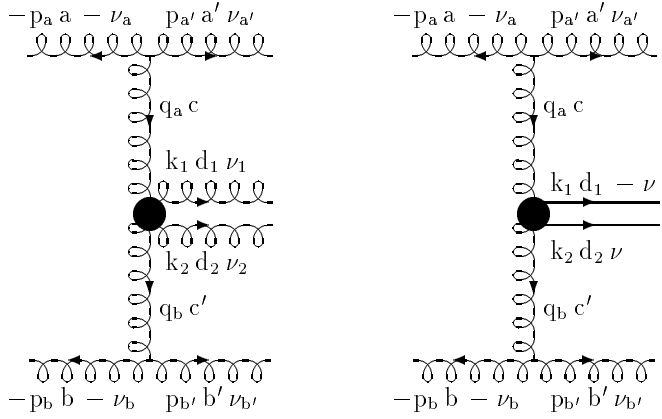}
\vspace*{-15cm}
\caption{Amplitudes for the production of four partons, with partons
$k_1$ and $k_2$ in the central-rapidity region.}
\label{fig:centr}
\end{figure}

We consider the production of four partons with momenta $p_{a'}$, 
$k_1$, $k_2$ and $p_{b'}$, in the scattering between two partons of momenta 
$p_a$ and $p_b$. We require that partons $k_1$ and $k_2$
have similar rapidity and are separated through large rapidity intervals
from the partons emitted in the forward-rapidity regions, with
all of them having comparable transverse momenta (Fig.\ref{fig:centr})
\begin{equation}
y'_A\gg y_1 \simeq y_2 \gg y'_B\,;\qquad |k_{1\perp}| \simeq |k_{2\perp}|
\simeq |p'_{A\perp}| \simeq |p'_{B\perp}|\, .\label{crreg}
\end{equation}
First we consider the amplitude for the scattering $g\, g\, \ra g\, g\, g\, g$
\cite{fl}-\cite{fl2} (Fig.\ref{fig:centr}a). Using eq.(\ref{two}) and the
amplitudes with three negative-helicity gluons \cite{bg}, we have 
\cite{ptlipnl}
\begin{eqnarray}
& & M^{g\, g}(-p_a,-\nu_a; p_{a'},\nu_{a'}; k_1,\nu_1; k_2,\nu_2; p_{b'},
\nu_{b'}; -p_b,-\nu_b) \label{centr}\\ & & = - 4\, g^4\, {\hat s 
\over |p_{a'\perp}|^2 |p_{b'\perp}|^2}\, C^{g\, g}_{-\nu_a\nu_{a'}}
(-p_a,p_{a'})\, C^{g\, g}_{-\nu_b\nu_{b'}}(-p_b,p_{b'}) \nonumber\\ & &
\times \left\{ A^{g\, g}_{\nu_1\nu_2}(k_1,k_2)
\left[ {\rm tr} \left( \lambda^a \lambda^{a'} \lambda^{d_1} \lambda^{d_2} 
\lambda^{b'} \lambda^b - \lambda^a \lambda^{a'} \lambda^{d_1} 
\lambda^{d_2} \lambda^b \lambda^{b'} \right. \right.\right. \nonumber\\ & & 
\left.\left. - \lambda^a \lambda^{d_1} \lambda^{d_2}
\lambda^{b'} \lambda^b \lambda^{a'} + \lambda^a \lambda^{d_1} \lambda^{d_2}
\lambda^b \lambda^{b'} \lambda^{a'} \right) 
+ {\rm traces}\, {\rm in}\, {\rm reverse}\, {\rm order} \right] 
- B^{g\, g}_{\nu_1\nu_2}(k_1,k_2) \nonumber\\ & &
\left. \times \left[ {\rm tr} \left( 
\lambda^a \lambda^{a'} \lambda^{d_1} \lambda^{b'} \lambda^b \lambda^{d_2}
- \lambda^a \lambda^{a'} \lambda^{d_1} \lambda^b \lambda^{b'} \lambda^{d_2} 
\right) + {\rm traces}\, {\rm in}\, {\rm reverse}\, {\rm order} \right]
+ \left(\begin{array}{c} k_1\leftrightarrow k_2\\ \nu_1\leftrightarrow \nu_2\\ 
d_1\leftrightarrow d_2 \end{array}\right) \right\}\, ,\nonumber
\end{eqnarray}
with the production vertices $C$ of gluons $p_{a'}$ and $p_{b'}$ determined by
eq.(\ref{centrc}), and the vertex for the production of gluons $k_1$ and 
$k_2$, $g^*\, g^* \rightarrow g\, g$, given by
\begin{eqnarray}
B^{g\, g}_{\nu_1\nu_2}(k_1,k_2) &=& A^{g\, g}_{\nu_1\nu_2}(k_1,k_2) + 
A^{g\, g}_{\nu_2\nu_1}(k_2,k_1)\,
,\label{cra}\\
A^{g\, g}_{++}(k_1,k_2) &=& 2\, {p_{a'\perp}^* p_{b'\perp}\over k_{1\perp}}
{1\over k_{2\perp} - k_{1\perp} {k_2^+\over k_1^+}}\, ,\nn\\
A^{g\, g}_{+-}(k_1,k_2) &=& -2\, {k_{1\perp}^* \over k_{1\perp}} 
\left\{ - {1\over \hat s_{12}} \left[{k_{2\perp}^2 |q_{a\perp}|^2 \over 
(k_1^-+k_2^-)k_2^+} +{k_{1\perp}^2 |q_{b\perp}|^2 \over (k_1^++k_2^+)k_1^-} 
+ {\hat t\, k_{1\perp}k_{2\perp}\over k_1^-k_2^+} \right]\right. \label{kosc}\\
&+& \left. {(q_{b\perp}+k_{2\perp})^2 \over \hat t} -
{q_{b\perp}+k_{2\perp} \over \hat s_{12}}
\left[{k_1^-+k_2^-\over k_1^-} k_{1\perp} - {k_1^++k_2^+\over k_2^+} 
k_{2\perp} \right]\right\} \nonumber\\
{\rm with} & & q_a = -(p_{a'} - p_a) \qquad q_b = p_{b'} - p_b \nonumber\\
{\rm and} & & \s_{12} = 2k_1\cdot k_2 \qquad \th \simeq - \left(
|q_{b\perp}+k_{2\perp}|^2 + k_1^-k_2^+ \right)\, .\nn 
\end{eqnarray}
Note that for equal helicities the vertices $A$ and $B$ are similar in
form to the respective vertices in eq.(\ref{treposb}) and (\ref{ab}),
because the helicity structure of the amplitudes they belong to is similar.
As in sect.~\ref{sec:fourone}, the vertex $A^{g\, g}_{\nu_1\nu_2}$ has
a collinear divergence as $\hat s_{12}\ra 0$, but the divergence
cancels out in the vertex $B^{g\, g}_{\nu_1\nu_2}$. In addition,
the amplitude (\ref{centr}) 
must not diverge more rapidly than $1/|q_{i\perp}|$ in the collinear regions
$|q_{i\perp}|\rightarrow 0$, with $i=a,b$, in order for the related 
cross section not to diverge more than logarithmically \cite{fl}.
Since eq.(\ref{centr}) has the poles $|q_{a\perp}|^2$ 
and $|q_{b\perp}|^2$, the $A$-vertex must be at least linear in $|q_{i\perp}|$,
\begin{equation}
\lim_{|q_{i\perp}|\rightarrow 0} A^{g\, g}_{\nu_1\nu_2}(k_1,k_2) =
O(|q_{i\perp}|)\, ,\label{aqlim}
\end{equation}
which is fulfilled by eq.(\ref{kosc}). Finally, in the soft limit 
$k_1\ra 0$ we obtain
\begin{equation}
\lim_{k_1\ra 0} A^{g\, g}_{+-}(k_1,k_2) = {q_{a\perp} q_{b\perp}^* k_{2\perp}
\over q_{a\perp}^* q_{b\perp} k_{2\perp}^*} A^{g\, g}_{++}(k_1,k_2)\, 
,\label{soft}
\end{equation}
which, when integrated over the phase space of gluon $k_1$, yields a
logarithmic infrared divergence.
Using the algebra (\ref{alg}) and eq.(\ref{cra}), and fixing $\th_i \simeq
- |q_{i\perp}|^2$ with $i = a, b$, the amplitude (\ref{centr}) may be
rewritten as
\begin{eqnarray}
& & M^{g\, g}(-p_a,-\nu_a; p_{a'},\nu_{a'}; k_1,\nu_1; k_2,\nu_2; p_{b'},
\nu_{b'}; -p_b,-\nu_b) \label{centrb}\\ & & = 2\,\s\, \left[ ig\, f^{aa'c}\,
C^{g\, g}_{-\nu_a\nu_{a'}}(-p_a,p_{a'})\right] {1\over \th_a}
\left\{ (ig)^2\, f^{cd_1e}\, f^{ed_2c'} A^{g\, g}_{\nu_1\nu_2}(k_1,k_2) +
\left(\begin{array}{c} k_1\leftrightarrow k_2\\ \nu_1\leftrightarrow \nu_2\\ 
d_1\leftrightarrow d_2 \end{array}\right) \right\} \nonumber\\ & & \times
{1\over \th_b} \left[ ig\, f^{bb'c'}\, C^{g\, g}_{-\nu_b\nu_{b'}}(-p_b,p_{b'})
\right]\, ,\nn
\end{eqnarray}
where we have enclosed the production vertex $g^*\, g^* \rightarrow g\, g$
of gluons $k_1$ and $k_2$ 
in curly brackets. In the multi-Regge limit $k_1^+\gg k_2^+$ it becomes
\begin{eqnarray}
\lim_{k_1^+\gg k_2^+} A^{g\, g}_{\nu_1\nu_2}(k_1,k_2) = C^g_{\nu_1}
(q_a,q_{12})\, {1\over \hat t_{12}} C^g_{\nu_2}(q_{12},q_b)\, ,\label{mrlim}
\end{eqnarray}
with $q_{12}$ the momentum of the gluon exchanged between $k_1$ and
$k_2$ and $\hat t_{12} \simeq - |q_{12\perp}|^2$;
thus the amplitude (\ref{centrb}) reduces to eq.(\ref{ntree}), with $n = 2$.
Finally, we note that fixing
\begin{eqnarray}
x &=& {k_1^+ \over k_1^++k_2^+}\, , \qquad \Delta_{\perp} = k_{1\perp} + 
k_{2\perp}\, , \qquad k_{2\perp} - k_{1\perp} {k_2^+\over k_1^+} = 
{x\Delta_{\perp} - k_{1\perp} \over x}\, ,\label{fldd}\\ \hat s_{12} &=& 
{|k_{1\perp} - x\Delta_{\perp}|^2 \over x(1-x)}\, ,\qquad \hat t =
- {|k_{1\perp} - x q_{a\perp}|^2 + x(1-x)|q_{a\perp}|^2 \over x}\, ,\nonumber
\end{eqnarray}
the A-vertex (\ref{kosc}) may be rewritten as \cite{fl2},
\begin{eqnarray}
A^{g\, g}_{++}(k_1,k_2) &=& -2\, {q_{a\perp}^* q_{b\perp}\over k_{1\perp}}
{x \over x\Delta_{\perp} - k_{1\perp} }\, ,\nn\\
A^{g\, g}_{+-}(k_1,k_2) &=& -2\, {k_{1\perp}^* \over k_{1\perp}} 
\left[ {(q_{b\perp}+k_{2\perp})^2 \over \th} + {x |q_{a\perp}|^2
k_{2\perp}^2 \over |\Delta_{\perp}|^2 (|k_{1\perp} 
- x\Delta_{\perp}|^2 + x(1-x) |\Delta_{\perp}|^2)} \right.
\label{smart}\\ &-& \left. { x(1-x) q_{a\perp} q_{b\perp}^* k_{2\perp} \over 
\Delta_{\perp}^* k_{1\perp}^* (k_{1\perp} - x \Delta_{\perp})} - 
{x q_{a\perp}^* q_{b\perp} k_{1\perp} \over |\Delta_{\perp}|^2 
(k_{1\perp}^* - x \Delta_{\perp}^*)} + {x q_{b\perp}^* (q_{b\perp}+k_{2\perp})
\over \Delta_{\perp}^* k_{1\perp}^*} \right]\, .\nonumber
\end{eqnarray}

Next, we consider the amplitude $g\, g \rightarrow g\, \bar{q}\, q\, g$,
with the production of a $\bar{q}\, q$ pair in the central-rapidity region
(Fig.\ref{fig:centr}b) \cite{fl2}, \cite{ptlipqq}. Using the amplitudes
with a $\bar{q}\, q$ pair and two negative-helicity gluons \cite{mpz}, and
the algebra (\ref{alg}) we obtain
\begin{eqnarray}
& & M^{\bar{q}\, q}(-p_a,-\nu_a; p_{a'},\nu_{a'}; k_1,-\nu; k_2,\nu; 
p_{b'},\nu_{b'}; -p_b,-\nu_b) \label{centrq}\\ & & = 4\, \hat s 
\left[ ig\, f^{aa'c}\, C^{g\, g}_{-\nu_a\nu_{a'}}(-p_a,p_{a'})\right] 
{1\over \th_a} \left\{g^2 \left[ (\lambda^{c'} \lambda^c)_{d_2\bar d_1}
A^{\bar{q}\, q}_{-\nu\nu}(k_1,k_2) - (\lambda^c \lambda^{c'})_{d_2\bar d_1}
A^{q\,\bar{q}}_{\nu-\nu}(k_2,k_1) \right] \right\} \nonumber\\ & & \times
{1\over \th_b} \left[ ig\, f^{bb'c'}\, C^{g\, g}_{-\nu_b\nu_{b'}}(-p_b,p_{b'})
\right]\, ,\nn
\end{eqnarray}
with $k_1$ the antiquark,
and the production vertex $g^*\, g^* \rightarrow \bar q\, q$ enclosed
in curly brackets. The vertex $A^{\bar{q}\, q}$ is
\begin{eqnarray}
A^{\bar{q}\, q}_{+-}(k_1,k_2) = &-& \sqrt{k_1^+\over k_2^+}
\left\{ {k_2^+ |q_{b\perp}|^2 \over (k_1^++k_2^+) \hat s_{12}} +
{k_2^- k_{2\perp} |q_{a\perp}|^2 \over k_{1\perp} (k_1^-+k_2^-) \hat s_{12}}
+ {k_2^+ k_{1\perp}^* (q_{b\perp}+k_{2\perp}) \over k_1^+ \hat t} \right.
\nonumber\\ &+& \left. {(q_{b\perp}+k_{2\perp}) [k_1^-k_2^+ - k_{1\perp}^* 
k_{2\perp} - (q_{b\perp}^*+k_{2\perp}^*) k_{2\perp}] \over k_{1\perp} 
\hat s_{12}} - {|k_{2\perp}|^2 \over \hat s_{12}} \right\}\, .\label{centqq}
\end{eqnarray}
As in the gluonic case, we note that the vertex $A^{\bar{q}\, q}_{+-}$
must be at least linear in $|q_{i\perp}|$,
\begin{equation}
\lim_{|q_{i\perp}|\rightarrow 0} A^{\bar{q}\, q}_{+-}(k_1,k_2) =
O(|q_{i\perp}|) \label{alim}
\end{equation}
with $i=a,b$, which is fulfilled by eq.(\ref{centqq}).
Using eq.(\ref{fldd}), the vertex $A^{\bar{q}\, q}_{+-}$
may be rewritten as \cite{fl2},
\begin{eqnarray}
A^{\bar{q}\, q}_{+-}(k_1,k_2) = &-& \sqrt{1-x\over x} \left[
{k_{1\perp}^* (q_{b\perp}+k_{2\perp}) \over \hat t} + {x |q_{a\perp}|^2
k_{1\perp}^* k_{2\perp} \over |\Delta_{\perp}|^2 (|k_{1\perp} 
- x\Delta_{\perp}|^2 + x(1-x) |\Delta_{\perp}|^2)} \right.
\nonumber\\ &-& \left. { x(1-x) q_{a\perp} q_{b\perp}^* \over \Delta_{\perp}^*
(k_{1\perp} - x \Delta_{\perp})} + {x q_{a\perp}^* q_{b\perp} k_{1\perp}^*
\over |\Delta_{\perp}|^2 (k_{1\perp}^* - x \Delta_{\perp}^*)} +
{x q_{b\perp}^* \over \Delta_{\perp}^*} \right]\, .\label{smartqq}
\end{eqnarray}
In the multi-Regge limits, $k_1^+\gg k_2^+$,
i.e. for $x \rightarrow 1$, or $k_2^+\gg k_1^+$, i.e. for $x \rightarrow 0$,
the vertex (\ref{smartqq}) vanishes. In addition, in the soft limit $k_1\ra 0$,
i.e. for $x \rightarrow 0$ {\sl and} $k_{1\perp} \rightarrow 0$, the vertex 
$A^{\bar{q}\, q}_{+-}$ (\ref{smartqq}) has a square-root divergence,
\begin{equation}
\lim_{k_1\ra 0}
A^{\bar{q}\, q}_{+-}(k_1,k_2) = {1\over \sqrt{x}} {x q_{a\perp} q_{b\perp}^* 
\over \Delta_{\perp}^* (k_{1\perp} - x \Delta_{\perp})}\, ,\label{softqq}
\end{equation}
which when integrated over the quark phase space does not yield any 
infrared divergence because soft quarks are infrared safe.
Finally, the amplitudes of Fig.~\ref{fig:centr} with a quark 
in the upper and/or in the lower line are obtained via the 
substitution (\ref{qlrag}).

\section{Conclusions}

In these proceedings we have shown how the helicity formalism may be
used in the high-energy limit to compute all the corrections to the 
tree-level FKL amplitudes induced by the corrections 
to the multi-Regge kinematics.
The building blocks of the ensuing amplitudes are the vertices which
describe the emission of two partons in the forward-rapidity region 
(sect.~4.1), or in the central-rapidity region (sect.~4.2). Once the
parton helicities are fixed the analytic form of the vertices simplifies
considerably (see ref.~\cite{fl} versus ref.~\cite{ptlipnl}, \cite{fl2}
and \cite{ptlipqq}).
The vertices for the emission in the central-rapidity region determine
the real NLL corrections to the BFKL equation \cite{fl2}, and the
vertices for the emission in the forward-rapidity region fix the boundary
conditions to it. 

\vspace{.5cm}

{\un {\sl Acknowledgements}} I wish to thank the organizers of Les Rencontres
de Physique de la Vallee d'Aoste for the warm hospitality and for the
support.

\end{document}